# Thermoelectric Cooperative Effect in Three-Terminal Elastic Transport through a Quantum Dot

**Jincheng Lu, Rongqian Wang, Yefeng Liu and Jian-Hua Jiang***

College of Physics, Optoelectronics and Energy, & Collaborative Innovation Center of Suzhou Nano Science and Technology, Soochow University, 1 Shizi Street, Suzhou 215006, China

* jianhuajiang@suda.edu.cn, joejhjiang@hotmail.com

**A B S T R A C T**

The energy efficiency and power of a three-terminal thermoelectric nanodevice are studied by considering elastic tunneling through a single quantum dot. Facilitated by the three-terminal geometry, the nanodevice is able to generate simultaneously two electrical powers by utilizing only one temperature bias. These two electrical powers can add up constructively or destructively, depending on their signs. It is demonstrated that the constructive addition leads to the enhancement of both energy efficiency and output power for various system parameters. In fact, such enhancement, dubbed as thermoelectric cooperative effect, can lead to maximum efficiency and power no less than when only one of the electrical power is harvested.

## 1. Introduction

Thermoelectric phenomena at nanoscales have attracted a lot of research interest because of fundamental physics and application impacts on renewable energy devices with high performance[1-16]. Theory and experiments have shown that nanostructured materials can have high thermoelectric efficiency and power[13,14,17-22]. Up till now, most of the theories for thermoelectricity is based on elastic (or quasi-elastic) transport theory, where energy-dependent conductivity is commonly involved[11,23-29]. In particularly, Mahan and Sofo proposed that the "best thermoelectrics" can be realized in narrow band conductors, where the thermopower and the electronic heat current are balanced and optmized to yield a high energy efficiency (characterized by a large thermoelectric figure of merit, $ZT$)[30]. However, recent studies show that if phonon parasitic heat conduction is taken into account the bandwidth of the carrier should be much enlarged and the figure of merit is significantly reduced[11,31]. These findings reveal the intrinsic entanglement of the Seebeck coefficient, electrical conductivity, and the heat conductivity, which may impede future improvement of thermoelectric performance.

Recently, to go beyond such an obstacle, the concept of inelastic thermoelectric transport is proposed[5,11,13,14,32,33]. A typical inelastic thermoelectric device involve three terminals (see Fig.1a): two electrodes (the source and the drain), and a boson bath (e.g., a phonon bat). The boson bath provides the energy (in the form of, e.g., phonons) to assist the inelastic transport between the source and the drain. This picture is essentially similar to a solar cell, but with much lower energy scales. In the situation of phonon-assisted hopping transport, the figure of merit is limited by the average frequency and bandwidth of the phonons involved in the inelastic transport[16]. High figure of merit can be achieved with large average frequency and small bandwidth[11,31], which do not conflict with electrical conductivity if the electron-phonon interaction is strong (e.g., electron-phonon interaction near the Debye frequency in ionic crystals)[5,34]. Thus high thermoelectric efficiency and power may be achieved without requiring narrow electronic bands. Such a paradigm also provides a three-terminal (or, in principle, multiterminal) geometry which enriches the manipulation of heat and electrical currents.

In this work we show that the three-terminal geometry can enable two thermoelectric powers induced by a single temperature bias. This effect, established by thermodynamic arguments, holds for both inelastic and elastic thermoelectric transport. By adopting a minimum quantum dot resonant tunneling model, we show that for elastic thermoelectric transport, a cooperative phenomenon emerge in the output electrical power: since the signs of the two thermoelectric powers can be controlled by the voltages, the two thermoelectric powers can add up constructively when they are both positive, leading to enhanced maximum output power. The maximum energy efficiency can be improved similarly. These thermoelectric cooperative effects hold for both elastic and inelastic thermoelectric transport, even in the linear-response regime (since neither output power nor energy efficiency is a linear function of the affinities). In fact, we can prove that the maximum efficiency and power are no less than their optimal values when only one electrical power is collected. Our results are consistent with recent studies on thermoelectric energy conversion in multiterminal devices.

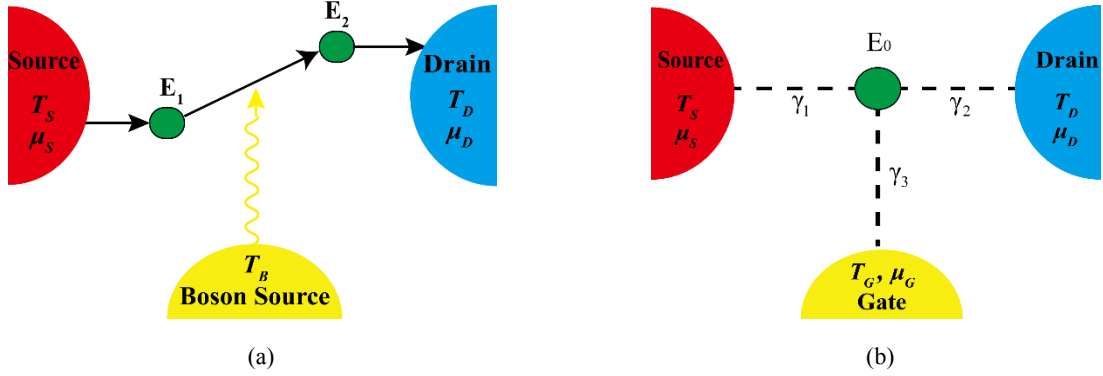

**Figure. 1.** (a) Sketch of three-terminal inelastic thermoelectric mesoscopic systems: two quantum dots with energy level $E_1$ and $E_2$ are connected to two electronic reservoirs, the source, the drain; the straight black arrows indicate the electronic currents, and the wavy yellow one represents the phonon heat current. (b) Sketch of the single dot model: a quantum dot with a single energy level $E_0$ is connected to three electronic reservoirs, the source, the drain and the gate. The electrochemical potential and temperature of the reservoir $i$ with $i = (S, D, G)$ are $\mu_i$ and $T_i$, respectively. The constants, $\gamma_1$, $\gamma_2$ and $\gamma_3$, represent the tunneling rates between the quantum dot and three rervoirs.

## 2. Linear Response For Three-Terminal Nanodevice with a Single Quantum dot

We consider a quantum system composed of a quantum dot (QD) coupled to three electrodes. In the linear-response regime, the charge and heat transport are governed by the Onsager matrix **M** via the relation

$$\begin{pmatrix} J_S^e \\ J_S^Q \\ J_G^e \\ J_G^Q \end{pmatrix} = \begin{pmatrix} M_{11} & M_{12} & M_{13} & M_{14} \\ M_{21} & M_{22} & M_{23} & M_{24} \\ M_{31} & M_{32} & M_{33} & M_{34} \\ M_{41} & M_{42} & M_{43} & M_{44} \end{pmatrix} \begin{pmatrix} A_S^\mu \\ A_S^T \\ A_G^\mu \\ A_G^T \end{pmatrix} \quad (1)$$

where $J_i^e(J_i^Q)$ represents the charge (heat) current entering the QD from the lead $i$, with $i = (S, G)$, see Fig.1(b). We define the affitinies $A_i^\mu = (\mu_i - \mu_D)/eT$ and $A_i^T = (T_i - T_D)/T^2$. Here $\mu_i$ and $T_i$ are the electrochemical potential and temperature, respectively, of the reservoir $i = (S, G)$, and $T$ is the equilibrium temperature. The Onsager coefficients $M_{ij}$ are symmetric $M_{ij} = M_{ji}$, and have been calculated in Ref.[35].

The coherent flow of charge and heat through a non-interacting ballistic conductor can be described by means of the Landauer–Büttiker formalism[4,36]. Under the assumption that all phase-breaking and dissipative processes take place in the reservoirs, the charge and thermal currents are

3 / 13

expressed in terms of the scattering properties of the system[7]. For example, in a generic multi-terminal configuration, the charge and heart currents flowing into the system from the $i$-th reservoir are:

$$J_i^e = \frac{e}{h}\sum_{j\neq i}\int_{-\infty}^{+\infty} dE \mathcal{T}_{ij}(E)[f_i(E) - f_j(E)] \tag{2}$$

$$J_i^Q = \frac{1}{h}\sum_{j\neq i}\int_{-\infty}^{+\infty} dE(E-\mu_i)\mathcal{T}_{ij}(E)[f_i(E) - f_j(E)] \tag{3}$$

where the sum over $j$ is intended over all but the $i$th reservoir, $h$ is the Planck constant, $\mathcal{T}_{ij}(E)$ is the transmission probability for a particle with energy $E$ to transit from the reservoir $i$ to reservoir $j$, and $f_i(E) = \{\exp[(E-\mu_i)/k_B T_i] + 1\}^{-1}$ is the Fermi distribution.

In this paper we focus on the situations where the heat current flowing out of the gate reservoir vanishes, i.e., $J_G^Q = 0$, via setting

$$A_G^T = -\frac{M_{41}A_S^\mu + M_{42}A_G^\mu + M_{43}A_S^T}{M_{44}} \tag{4}$$

By substituting Eq. (4) into Eq. (1) we obtain the following transport equation:

$$\begin{pmatrix} J_S^e \\ J_G^e \\ J_S^Q \end{pmatrix} = \begin{pmatrix} M'_{11} & M'_{12} & M'_{13} \\ M'_{21} & M'_{22} & M'_{23} \\ M'_{31} & M'_{32} & M'_{33} \end{pmatrix} \begin{pmatrix} A_S^\mu \\ A_G^\mu \\ A_S^T \end{pmatrix} \tag{5}$$

The transport coefficients can be written in a simplified way as:

$$M'_{11} = M_0\gamma_1\left(\gamma_2 + \gamma_3 - \alpha\frac{\gamma_1\gamma_2}{\gamma_1 + \gamma_3}\right)e^2$$

$$M'_{12} = M'_{21} = M_0(-\gamma_1\gamma_2)(1-\alpha)e^2$$

$$M'_{13} = M'_{31} = M_1\gamma_1\left(\gamma_2 + \gamma_3 - \frac{\gamma_1\gamma_2}{\gamma_1 + \gamma_3}\right)e$$

$$M'_{22} = M_0\gamma_2(\gamma_1 + \gamma_3)(1-\alpha)e^2$$

$$M'_{23} = M'_{32} = 0$$

$$M'_{33} = M_2\gamma_1\left(\gamma_2 + \gamma_3 - \frac{\gamma_1\gamma_2}{\gamma_1+\gamma_3}\right) \tag{6}$$

where $\alpha \equiv M_1^2/(M_0 M_2)$ with $M_n \equiv \frac{T}{h}\int dE(-\frac{\partial f}{\partial E})(E-\mu)^n\mathcal{T}$ and $\mathcal{T} = \left[(E-E_0)^2 + \left(\frac{\gamma_1+\gamma_2+\gamma_3}{2}\right)^2\right]^{-1}$. Here $f(E) = \{\exp[(E-\mu)/k_B T] + 1\}^{-1}$ is the equilibrium distribution. Importantly, from the Cauchy-Schwarz inequality,

$$0 < \alpha < 1 . \tag{7}$$

$M'_{11}$ and $M'_{22}$ represent the electrical conductance, $M'_{12}=M'_{21}$ stand for the off-diagonal electrical conductance, $M'_{13} = M'_{31}$ and $M'_{23} = M'_{32}$ are the Seebeck coefficients, and $M'_{33}$ is the thermal conductance.

The total entropy production of the system in the linear response regime is written as

$$\frac{dS}{dt} = \frac{1}{T}\left(J_S^e A_S^\mu + J_G^e A_G^\mu + J_S^Q A_S^T\right) \tag{8}$$

The second law of thermodynamics $\frac{dS}{dt} \geq 0$, requires that[2]

$$M'_{11}M'_{22} \geq {M'_{12}}^2, \; M'_{11}M'_{33} \geq {M'_{13}}^2, \; M'_{22}M'_{33} \geq {M'_{23}}^2 \tag{9}$$

as well as that the determinant of the $3 \times 3$ transport matrix in Eq. (5) to be non-negative. These requirements are all satisfied for the transport coefficients given in Eq. (6).

## 3. Cooperative Effect: A Geometric Interpretation

The two electrical affinities can be parametrized as

$$A_S^\mu = A^\mu \cos\theta, \; A_G^\mu = A^\mu \sin\theta \tag{10}$$

where $A^\mu = \sqrt{(A_S^\mu)^2 + (A_G^\mu)^2}$ is the total "magnitude" of the electrical affinities. To facilitate the discussion, we defined the effective electrical conductance as a function of the angle $\theta$

$$G_{eff}(\theta) = M'_{11}\cos^2\theta + 2M'_{12}\sin\theta\cos\theta + M'_{22}\sin^2\theta, \tag{11}$$

while the effective thermoelectric coefficient and the thermal conductance are,

$$L_{eff}(\theta) = M'_{13}\cos\theta, \qquad K = M'_{33}, \tag{12}$$

respectively. Each angle $\theta$ corresponds to a particular configuration between the two electrical affinities. By tuning $\theta$ we can obtain various configurations to explore the interference between the two thermoelectric effects.

The energy efficiency of the thermoelectric engine is given by[1,37]

$$\eta = -\frac{W}{J_S^Q} = -\frac{(J_S^e A_S^\mu + J_G^e A_G^\mu)T}{J_S^Q} \leq \eta_{max} = \eta_C \frac{\sqrt{1+ZT}-1}{\sqrt{1+ZT}+1}. \tag{13}$$

The Carnot efficiency is $\eta_C = \frac{(T_S-T_D)}{T_S}$. We find that the figure of merit is given as

$$ZT(\theta) = \frac{L_{eff}^2(\theta)}{G_{eff}(\theta)K - L_{eff}^2(\theta)} \tag{14}$$

Upon optimizing the output electrical power for a given $\theta$, we obtain[38]

$$W(\theta) = \frac{1}{4}P(\theta)(A_S^T)^2 \tag{15}$$

where the power factor is

$$P(\theta) = \frac{{M'_{13}}^2 \cos^2\theta}{M'_{11}\cos^2\theta + 2M'_{12}\sin\theta\cos\theta + M'_{22}\sin^2\theta} = \frac{L_{eff}^2(\theta)}{G_{eff}(\theta)} \tag{16}$$

Now we shall denote the thermoelectric energy conversion associated with $A_S^\mu$ as the "longitudinal thermoelectric effect". For that associated with $A_G^\mu$ we shall call it "transverse thermoelectric effect", since the former corresponds to the situation with parallel temperature and voltage gradient, whereas the latter corresponds to the voltage generated by the transverse temperature gradient. Without breaking the time-reversal symmetry or involving inelastic processes, it is not surprising that the transverse thermoelectric coefficient $M'_{23}$ vanishes.

When $\theta = 0$ or $\pi$, Eqs. (12) and (13) give the well-known figure of merit and power factor for the longitudinal thermoelectric effect[30]

$$Z_l T = \frac{{M'_{13}}^2}{M'_{11}M'_{33} - {M'_{13}}^2}, \qquad P_l = \frac{{M'_{13}}^2}{M'_{11}} \tag{17}$$

The transverse thermoelectric figure of merit and power factor, i.e., $\theta = \pi/2$ or $3\pi/2$, are given by

$$Z_t T = \frac{{M'_{23}}^2}{M'_{33}L'_{22} - {M'_{23}}^2} = 0, \qquad P_t = 0 \tag{18}$$

One can tune $\theta$ to maximize the figure of merit $ZT$ and electrical power $P$ which is achieved at

$$\frac{\partial ZT}{\partial \theta} = 0, \qquad \frac{\partial P}{\partial \theta} = 0. \tag{19}$$

After some algebraic calculation, we find that the maximum figure of merit and electrical power factor are

$$Z_m T = \frac{{M'_{13}}^2}{M'_{11}M'_{33} - {M'_{13}}^2 - {M'_{12}}^2 M'_{33}/M'_{22}} > \frac{{M'_{13}}^2}{M'_{11}M'_{33} - {M'_{13}}^2} = Z_l T, \tag{20}$$

$$P_m = \frac{{M'_{13}}^2}{M'_{11} - {M'_{12}}^2/M'_{22}} > \frac{{M'_{13}}^2}{M'_{11}} = P_l, \tag{21}$$

respectively.

At first sight, the above result is puzzling: the transverse thermoelectric effect vanishes, yet both the power factor and the figure of merit can be improved when both the longitudinal and the transverse thermoelectric effects are exploited to generate electrical power. In the remaining of this section we shall explain why such phenomenon is reasonable and show how the two electrical powers cooperative with each other.

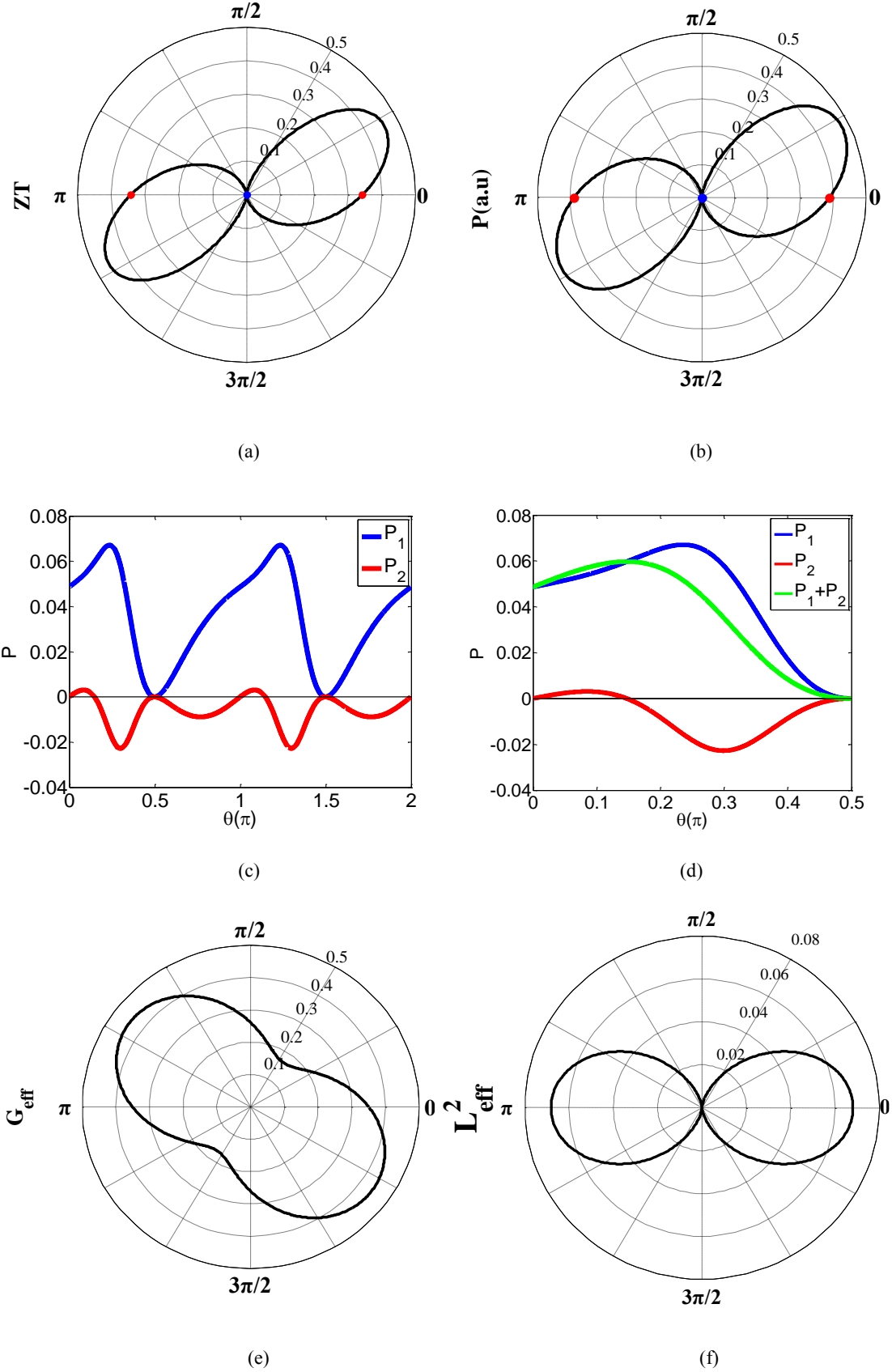

**Fig. 2.** (a) Polar plot of $ZT(\theta)$ vs $\theta$. (b) Polar plot of $P(\theta)$ (in arbitrary unit) vs $\theta$. At $\theta = 0$ or $\pi$, $ZT$ and $P$ recover the values for the longitudinal thermoelectric effect (red dots), while at $\theta = \pi/2$ or $3\pi/2$ they go back to those of the transverse thermoelectric effect (blue dot). (c) and (d) plot of the power $P_1 = -J_S^e A_S^\mu T$ and $P_2 = -J_G^e A_G^\mu T$ as a function of $\theta$ at the maximum power condition. (d) Focus of the $0$---$\pi/2$ region for both $P_1$, $P_2$, and their sum $P_1 + P_2$. (e) The effective conductance $G_{eff}(\theta)$ and (f) the effective thermoelectric coefficient $L_{eff}^2(\theta)$ as functions of $\theta$. The parameters are $\gamma_1 = \gamma_2 = \gamma_3 = k_B T$ and $E_0 = 2k_B T$.

To show a global view of how the two thermoelectric powers cooperate with each other, we plot in Fig. 2(a) and 2(b) the figure of merit $ZT$ and the power factor $P$ versus the angle $\theta$ in a polar plot for a set of physical parameters specified in the caption. Remarkably, the figure of merit $ZT$ can be greater than both $Z_l T$ and $Z_t T$ for $0 < \theta < \frac{\pi}{4}$. In the same region, the power factor $P$ can also be greater than both $P_l$ and $P_t$. In fact the power factor and the figure of merit follow the same trend in our system because $ZT(\theta) = \frac{P(\theta)}{K - P(\theta)}$.

The enhancements of the power factor and the figure of merit signify the cooperative thermoelectric effect, i.e., the two thermoelectric powers, $P_1 = -J_S^e A_S^\mu T$ and $P_2 = -J_G^e A_G^\mu T$, interfere constructively with each other. As shown in Figs.2(c) and 2(d), the two powers normally have opposite signs, unless for $0 < \theta < \frac{\pi}{4}$. In this region, both $P_1$ and $P_2$ are positive, leading to enhanced output power and energy efficiency [see Figs. 2(b) and 2(d)]. Physically, the cooperative effect originates from the coupling between the two charge transport channel, namely the term described by $M'_{12}$. Note that $M'_{12}$ is negative, and hence it can transfer electrical power between the two channels. In fact, it is responsible for the positiveness of $P_2$, since $M'_{22} > 0$ is related to Joule heat (i.e., negative contribution to the output power) and $M'_{23} = 0$. Moreover, even when $P_2$ becomes negative, the negative $M'_{12}$ leads to an increase of $P_1$ due to the increase of $A_G^\mu$, despite $A_S^\mu$ decreases with increasing $\theta$ in Fig. 2(d). We shall show below that such electrical power transfer reduces the total Joule heat and hence leads to increase of useful energy output and improved power factor.

The cooperative effect can also be manifested as the reduction of the Joule heat, which is related to the effective electrical conductance as $G_{eff}(A^\mu)^2 T^2$. As shown in Fig. 2(e) the effective conductance $G_{eff}$ is reduced (hence the total Joule heat is reduced) for $0 < \theta < \frac{\pi}{4}$. This is the reason that the power factor is enhanced despite that the effective thermoelectric coefficient $|L_{eff}(\theta)|$ is reduced, since the total power factor is given by $P(\theta) = \frac{L_{eff}^2(\theta)}{G_{eff}(\theta)}$. The improvement of the power factor then leads to the improvement of the figure of merit since $ZT(\theta) = \frac{P(\theta)}{K-P(\theta)}$.

**4. Cooperative thermoelectric effects for various configurations**

We now study the enhancement of the figure of merit and the power factor for various configurations. Since the transverse figure of merit and power factor vanishes. We shall compare the maximum figure of merit and power factor with the longitudinal figure of merit and power factor. We first consider the situation with $\gamma_1 = \gamma_2 = \gamma_3 = \gamma$. Using Eqs. (6) and (20), we obtain

$$Z_m T = \frac{\alpha}{1-\alpha} \tag{22}$$

where $\alpha = M_1^2/(M_0 M_2)$ with $M_n = \frac{T}{h} \int dE (-\frac{\partial f}{\partial E})(E-\mu)^n \mathcal{T}, (n=0,1,2)$ and $\mathcal{T} = [(E-E_0)^2 + 9\gamma^2/4]^{-1}$. Obviously the above figure of merit is always greater than the longitudinal thermoelectric figure of merit $Z_l T = \frac{3\alpha}{4(1-\alpha)}$.

The power factor of the total maxmium output power is

$$P_m = \frac{3}{2}\frac{M_1^2}{M_0}\gamma^2 \tag{23}$$

which is always greater than the longitudinal power factor, $P_l = \frac{9M_1^2 \gamma^2}{2M_0(4-\alpha)}$.

Comparing the figure of merit and power factors discussed in above section, we find that

$$\frac{Z_m T}{Z_l T} = \frac{4}{3} > 1 \tag{24a}$$

$$\frac{P_m}{P_l} = \frac{4-\alpha}{3} > 1 \tag{24b}$$

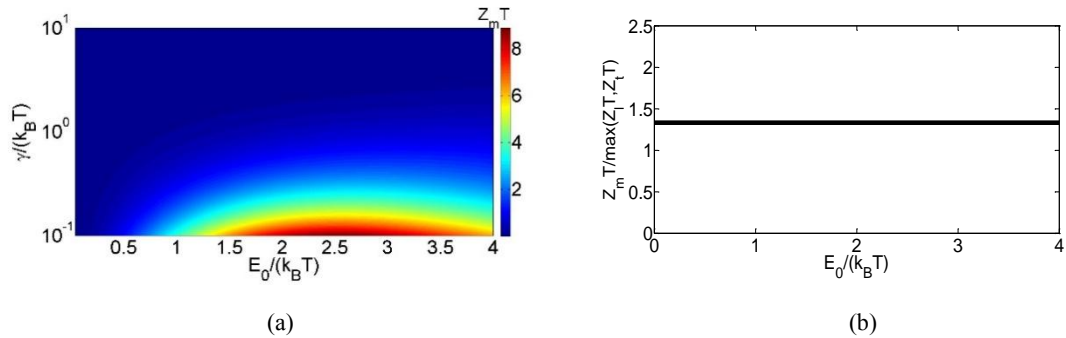

(a)      (b)

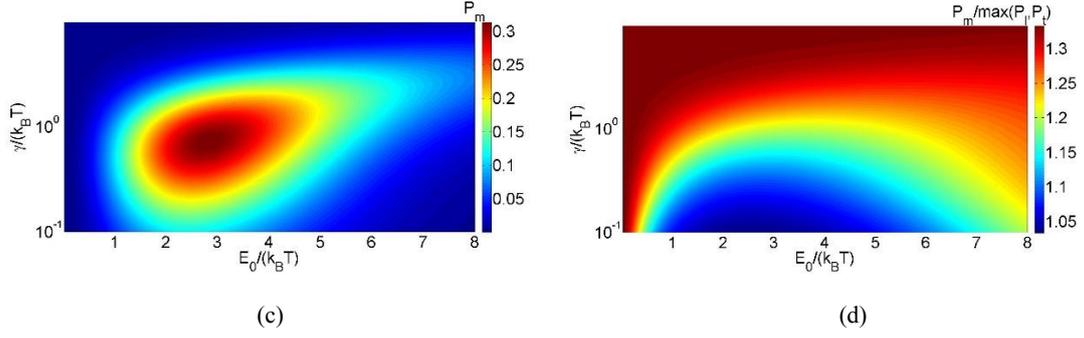

(c)                  (d)

**Fig. 3.** (a) The cooperative figure of merit $Z_m T$ and (b) the enhancement of figure of merit due to cooperative effect $Z_m T/\max(Z_l T, Z_t T)$ as functions of $E_0$ and $\gamma$, respectively. (c) The cooperative power $P_m$ as a function of $E_0$ and $\gamma$. (d) The enhancement of power factor due to cooperative effect, $P_m/\max(P_l, P_t)$, as a function of $E_0$ and $\gamma$.

Fig. 3(a) and (b) indicate that the figure of merit $Z_m T$ and $Z_m T/\max(Z_l T, Z_t T)$ as a function of the QD $E_0$, where we have set $\gamma_1 = \gamma_2 = \gamma_3 = \gamma$. The figure of merit $Z_m T$ is large when $\gamma$ is small, particularly for $E_0 \approx 2.5 k_B T$, in consistent with Mahan and Sofo[30]. The enhancement of the figure of merit is indeed a constant value $\frac{Z_m T}{\max(Z_l T, Z_t T)} = \frac{4}{3}$. As shown in Fig. 3(c) that the power factor $P_m$ induced by the cooperative effect is large for $E_0 \approx 3 k_B T$ and $\gamma \approx k_B T$. Fig. 3(d) shows the enhancement is large unless for the situations with both small $\gamma$ and $k_B T < E_0 < 6 k_B T$.

Next, we study the situation with $\gamma_1 = \gamma_2 \neq \gamma_3$. The maximum figure of merit remains the same, while the longitudinal figure of merit becomes $Z_l T = \frac{\alpha \gamma_3 (\gamma_3 + 2\gamma_1)}{(1-\alpha)(\gamma_1 + \gamma_3)^2}$.

The maxmium output power factor is

$$P_m = \frac{M_1^2 \gamma_1 \gamma_3 (\gamma_3 + 2\gamma_1)}{M_0 (\gamma_1 + \gamma_3)} \tag{25}$$

Comparing the figure of merit and power factors discussed in above section, we find that

$$\frac{Z_m T}{Z_l T} = \frac{(\gamma_1 + \gamma_3)^2}{\gamma_3 (\gamma_3 + 2\gamma_1)} > 1 \tag{26a}$$

$$\frac{P_m}{P_l} = \frac{(\gamma_1 + \gamma_3)^2 - \alpha \gamma_1^2}{\gamma_3 (2\gamma_1 + \gamma_3)} > 1 \tag{26b}$$

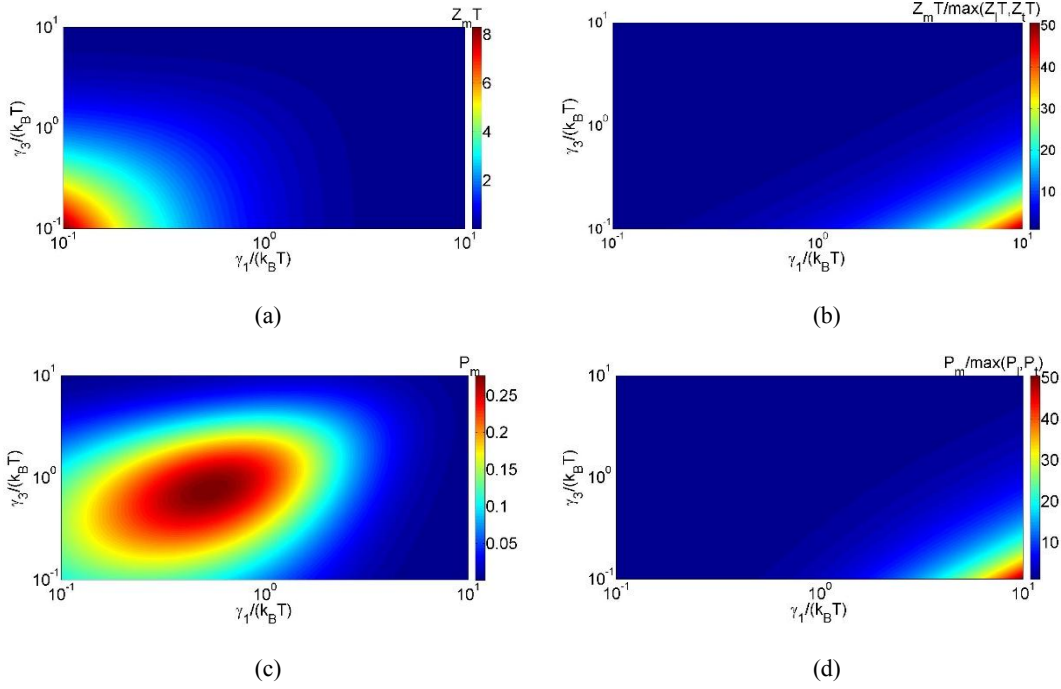

(a) (b)

(c) (d)

**Fig. 4.** (a) The cooperative figure of merit $Z_mT$ as a function of $\gamma_1$ and $\gamma_3$. (b) The enhancement of figure of merit due to cooperative effect, $Z_mT/\max(Z_lT, Z_tT)$, as a function of $\gamma_1$ and $\gamma_3$. (c) The electrical power $P_m$ as a function of $\gamma_1$ and $\gamma_3$. (d) The enhancement of figure of merit due to cooperative effect, $P_m/\max(P_l, P_t)$, as a function of $\gamma_1$ and $\gamma_3$. $E_0 = 2k_BT$.

Fig. 4(a) shows that the maximum figure of merit is large when both $\gamma_1$ and $\gamma_3$ are small, which is consistent with the picture that smaller linewidth leads to larger figure of merit. Fig. 4(b) indicates that the enhancement is large when $\gamma_3$ is small but $\gamma_1$ is large. From Eq. (25) this is because the enhancement increases with the ratio $\gamma_1/\gamma_3$. The power factor $P_m$, however, is large for $\gamma_1$ and $\gamma_3$ around $1k_BT$, as shown in Fig. 4(c). Finally, Fig. 4(d) shows that the enhancement of the power factor follows the same trend as that of the figure of merit, i.e., the enhancement is significant for large $\gamma_1/\gamma_3$.

Lastly, we make a comparison between the figure of merit and power factor at the general case $\gamma_1 \neq \gamma_2 \neq \gamma_3$ and find that

$$\frac{Z_mT}{Z_lT} = \frac{(\gamma_1+\gamma_3)(\gamma_2+\gamma_3)-\alpha\gamma_1\gamma_2-\alpha\gamma_3(\gamma_1+\gamma_2+\gamma_3)}{\gamma_3(1-\alpha)(\gamma_1+\gamma_2+\gamma_3)} \qquad (27a)$$

$$\frac{P_m}{P_l} = \frac{(\gamma_1+\gamma_3)(\gamma_2+\gamma_3)-\alpha\gamma_1\gamma_2}{\gamma_3(\gamma_1+\gamma_2+\gamma_3)}. \qquad (27b)$$

One can show that the above two enhancement ratios are always greater than one because of the inequality (7). Therefore, the thermoelectric figure of merit and the power factor can always be enhanced by the cooperative effect.

**5. Conclusion and Discussions**

We have shown that cooperative effects can be a potentially useful tool in improving the energy efficiency and output power in multi-terminal mesoscopic thermoelectric transport. The three-terminal geometry enables two charge currents coupled with one heat current, which can be exploited to generate two electrical powers simultaneously via a single temperature bias. Depending on their signs, the two electrical powers can lead to constructive or destructive addition. We show that the constructive addition leads to the cooperative thermoelectric effect that enhances both the output power and the energy efficiency. Through tuning the QD level and its coupling with the reservoirs, we can observe remarkable enhancement of the figure of merit and/or the power factor thanks to the thermoelectric cooperative effect. We have also shown how to understand the cooperative effect via the reduction of the total Joule heating. Our findings provide new opportunities for improving the thermoelectric performances of nanostructured materials.

**Acknowledgment**

That work is supported by the National Natural Science Foundation of China (no.11675116) and the Soochow University. JHJ thanks Prof. Yoseph Imry for inspiring discussions. JL thanks Tata Institute of Fundamental Research for hospitality.